\documentclass[aps,twocolumn,showpacs,superscriptaddress,floatfix]{revtex4}
\usepackage{graphicx}

\begin{document}

\title{Controllable Adiabatic  Manipulation of the Qubit State}
\author{G.P. Berman}
\affiliation{Theoretical Division, T-13 and CNLS, MS B213, Los
Alamos National Laboratory, Los Alamos, NM 87545}

\author{A.R. Bishop}
\affiliation{Theory, Simulation \& Computation Directorate, Los
Alamos National Laboratory, MS B210, Los Alamos, New Mexico, 87545}

\author{F. Borgonovi}
\affiliation{Dipartimento di Matematica e Fisica, Universit\'a Cattolica,
via Musei 41, 25121 Brescia, Italy}
\affiliation{INFN, Sezione di Pavia, Italy}

\author{V.I. Tsifrinovich}
\affiliation{Department of Physics, Polytechnic University, Six
Metrotech Center, Brooklyn, New York 11201}

\date{\today}

\begin{abstract}

{We propose a scheme which implements a controllable change of the
state of the target spin qubit in such a way that both the control
and the target spin qubits remain in their ground states. The
interaction between the two spins is mediated by an auxiliary spin,
which can transfer to its excited state. Our scheme suggests a
possible relationship between the gate and adiabatic quantum
computation. }

\end{abstract}


\maketitle

Quantum annealing and adiabatic quantum computation have attracted
much attention recently.(See, for example, \cite{uno}-\cite{tre}.)
Unlike the traditional (gate) quantum computer,  the adiabatic
quantum computer is based on a slow change of the Hamiltonian
describing the quantum system. The basic idea behind adiabatic
quantum computation is the following: in order to find a complicated
ground state of an Ising system in a longitudinal external magnetic
field, one starts from the simple ground state of the
non-interacting spins in an external transverse  magnetic field.
Then, one adiabatically changes the initial Hamiltonian to the Ising
one, so that, finally, the system exhibits the complicated ground
state of the Ising Hamiltonian. In the process of evolution, the
adiabatic quantum computer remains in its ground state. This
approach promises to solve important combinatorial and graph theory
NP-hard problems. One example is the MAX CLIQUE problem. In  graph
theory, a clique is a subset of vertices, such that every pair of
vertices is connected by an edge. In some cases, the MAX CLIQUE
problem is equivalent to finding  the ground state of the Ising
system \cite{due}. Adiabatic quantum computation has been
implemented recently by D-wave Systems Inc. using superconducting
flux qubits, whose evolution can be described by effective Ising
Hamiltonian with the controllable interaction constants
\cite{cinque}.

A ``traditional'' quantum computer is based on quantum logic gates,
which change the state of a quantum system. (See, for example,
\cite{quattro}.) Before and after the action of the gates the
quantum system is described by the same Hamiltonian. At  first
sight, an adiabatic quantum computer is completely different from a
gate quantum computer. Indeed, while a  gate quantum computer
utilizes quantum superposition, entanglement and interference in
order to ``sample'' all possible ``numbers'',   an adiabatic quantum
computer utilizes quantum tunneling in order to approach the true
ground state.

In this paper we investigate a possible bridge between the adiabatic
and gate approaches to quantum computation. Namely, we set the
simplest problem: how can one change the state of a target spin
qubit conditional on the state of a control spin qubit if both the
control and the target qubits remain in their ground states? One way
to achieve this objective is to use an auxiliary spin, which
mediates the interaction between the control and target spin qubits.
As an example, the first control qubit (an electron spin
$\vec{S}_1$) experiences a large local magnetic field $\vec{B}_1$
and always points opposite to this field (as the electron
gyromagnetic ratio is negative). An auxiliary spin $\vec{S}_{3}$
experiences a local magnetic field $\vec{B}_{3}$ and a {\it rf}
rotating field $\vec{B}_{rf}$.
 It also interacts with both the control spin $\vec{S}_{1}$ and
the target spin $\vec{S}_{2}$ (e. g. a ferromagnetic exchange
interaction with constants $J_{13}$ and $J_{23})$. The effective
exchange field $J_{13} \ \vec{S}_{3}$ (in frequency units) acting on
the spin $\vec{S}_{1}$ must be much smaller than $\vec{B}_{1}$ (in
the same units). The effective exchange fields $J_{13}\ \vec{S}_{1}$
and  $J_{23}\ \vec{S} _{2}$ must be much smaller than $\vec{B}_{3}$.
In this case one can use the resonant {\it rf} field $\vec{B}_{rf}$
on the auxiliary spin in order to manipulate its direction,
conditional on the direction of the control spin. The target spin
$\vec{S}_{2}$ experiences only the exchange field $J_{23}
\vec{S}_{3}\ $ produced by the auxiliary spin and should evolve
adiabatically, changing its direction together with the direction of
the auxiliary spin. Thus, the target spin will change its direction
conditional on the direction of the control spin remaining in the
ground state. The only spin which does not remain in the ground
state is the auxiliary one.

The greatest challenge in this proposed scheme is associated with
the adiabatic motion of the target spin. Indeed, the exchange field
produced by the target spin on the auxiliary spin must be small
compared to the exchange field produced by the control spin.
However, in this case the exchange field $J_{23} \vec{S}_{3 }$
produced by the auxiliary spin on the target one (which determines
the Larmor frequency of precession of the target spin) is small
compared to the field $\vec{B}_{3}$ acting on the auxiliary spin.
The magnetic field $\vec{B}_{3}$ determines the frequency of the
Larmor precession of the auxiliary spin and, correspondingly, the
frequency of oscillations of the exchange field on the target spin.
Thus, the adiabatic condition is violated for the target spin: the
frequency of the oscillation of the exchange field on the target
spin (which is determined by the field $\vec{B}_{3})$ will be
greater than the Larmor frequency of the target spin (which is equal
to $J_{23}\vec{S}_{3})$.

\begin{figure}
\includegraphics[angle=0,scale=0.33]{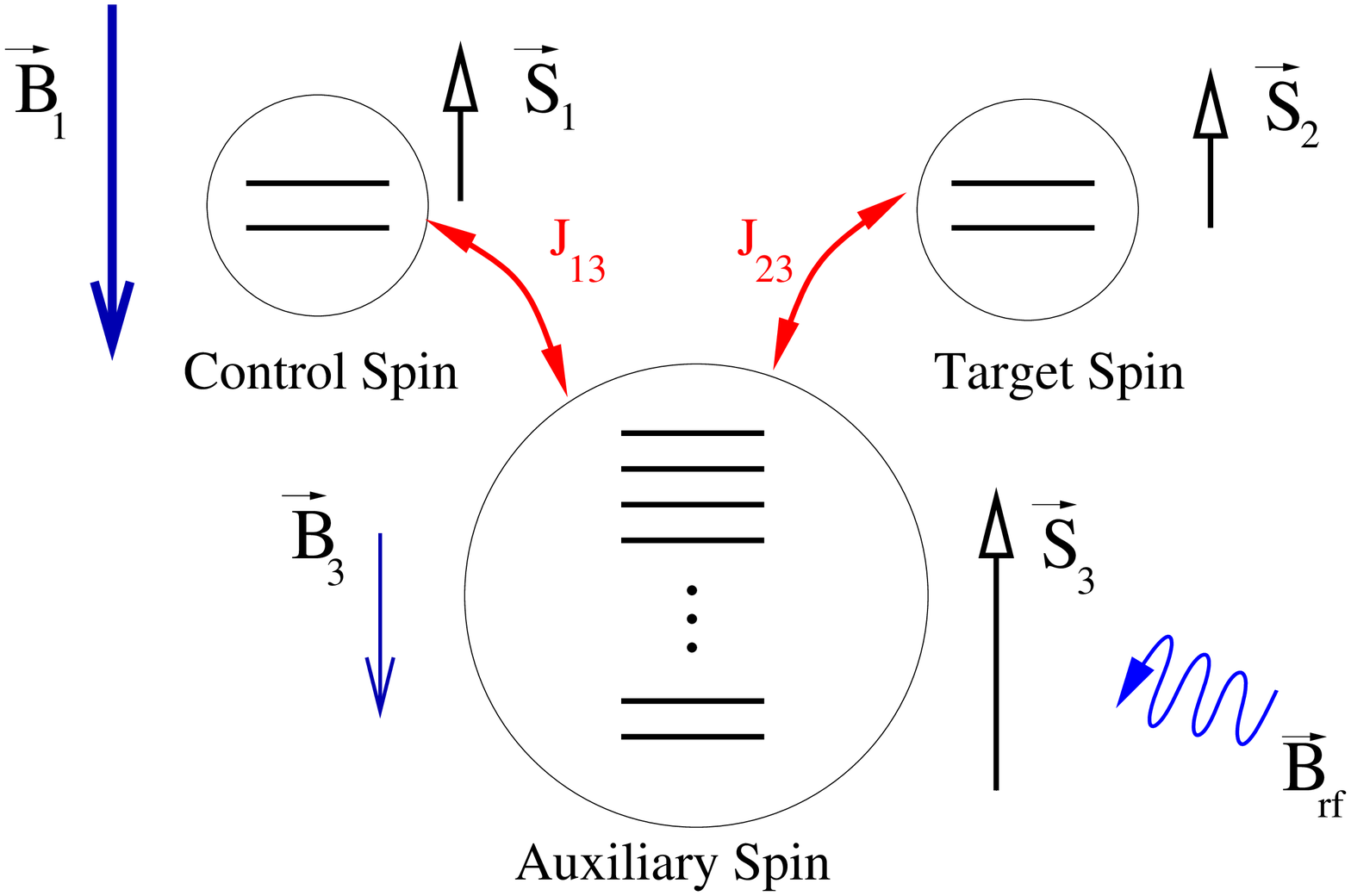}
\caption{Three-spin system.} \label{figuno}
\end{figure}

In order to avoid this problem we propose using an auxiliary spin $S
\gg 1/2$ . In this case one can increase the Larmor frequency of the
target spin without significantly changing other parameters except
for the field $\vec{B}_{1}$, which must remain much greater than the
exchange field $J_{13} \vec {S}_{3 }  $ on the control spin. Below
we describe our computer simulation with our proposed model. A
schematic of the spin system is shown in Fig.~\ref{figuno}.

The Hamiltonian of the system is

\begin{equation}
\begin{array}{lll}
H &=  \vec{ B}_{1} \vec{S}_{1}  +
\vec{B}_{3} \vec{S}_{3} - J_{13} \vec{S}_{1} \vec{S}_{3
} - J_{23} \vec{S}_{2} \vec{S}_{3} &+
\\
&&\\
& \frac{1}{2}  B_{rf} {\{}e^{i\omega t} { S}_{3}^+ +
e^{ - i\omega t} {S} _{3}^- {\}}.\\
\label{eq1}
\end{array}
\end{equation}

Here we set $\gamma $ = 1 ($\gamma $ is the magnitude of the
electron gyromagnetic ratio) and $\hbar = 1$. The {\it rf} field
rotates in the clockwise direction.

The parameters chosen for the computer simulations are:

\begin{equation}
\begin{array}{lll}
J_{23} &= 2, \quad J_{13} = 20, \quad B_{3} = 25, \quad
B_{1} = 2500, \quad B_{rf} = 3, &\\
&&\\
S_{1} &= S_{2} = 1/2, \quad  S_{3} = 51/2, \quad \omega  = 15.
\end{array}
\label{numbers}
\end{equation}

The duration $\tau $ of the action of the {\it rf}
 field corresponds to the $\pi $
pulse:

\begin{equation}
B_{rf} \tau =\pi.
\label{eq3}
\end{equation}

The initial conditions describe an auxiliary spin $S_{3}$ and the
target spin $S_{2}$ pointing ``up'' in the positive $z$-direction,
while the control spin $S_{1}$ may point ``up'' (as shown in
Fig.~\ref{figuno}) or ``down'' (not shown in Fig.~\ref{figuno}).

For our chosen parameters, the local magnetic field on the control
spin $B_{1} = 2500$ is much greater than the exchange field $J_{13}
S _{3}$ = 510. So, one can expect that the direction of the control
spin is determined by the local field. Next, for the auxiliary spin
the local magnetic field $B_{3}$ = 25 is more than twice the
exchange field $J_{13} S_{1}$ = 10  produced by the control spin and
much greater than the exchange field $J_{23} S _{2}$ = 1 produced by
the target spin. The  {\it rf}  field $B_{rf}$ =3 is greater than
the exchange field produced by the target spin but smaller than the
exchange field produced by the control spin. Thus, one can expect
that the action of the {\it rf}  pulse  on the auxiliary spin
depends on the direction of the control spin and does not depend on
the direction of  the target spin. If we ignore the influence of the
target spin,  the resonant frequency of the Larmor precession of the
auxiliary spin is ``35'' for the control spin pointing ``up'' and
``15'' for the control spin pointing ``down''. With $\omega $ = 15
the auxiliary spin changes its direction only if the control spin
points ``down''. Finally, the exchange field on the target spin
$J_{23} S_{3}$ = 51 is much greater than the expected frequency of
its oscillations: the transverse component of the exchange field is
expected to oscillate with the frequency $\omega $ = 15, and the
$z$-component with frequency $B_{rf}$ = 3. (The magnetic field has
the same unit as the frequency, as we set $\gamma $ = 1.) Thus, our
scheme is expected to implement a controlled change of the target
spin state.

\begin{figure}[t]
\includegraphics[angle=0,scale=0.36]{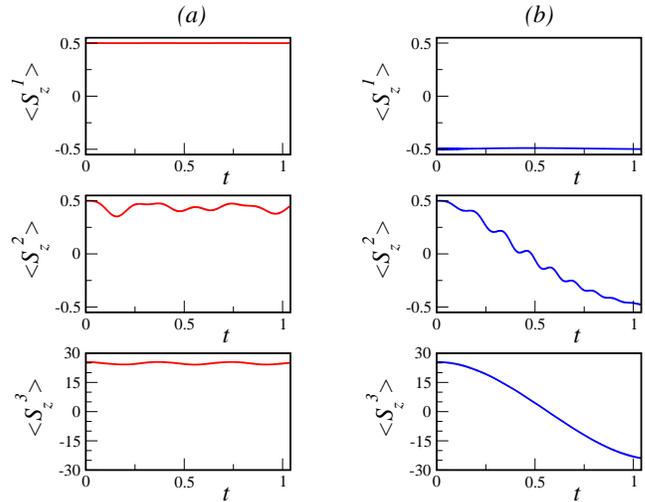}
\caption{
 Average spin evolution: a) -- for the control spin pointing ``up'',
and b) -- for the control spin pointing ``down''.
}
\label{figdue}
\end{figure}

The Schr\"odinger equation for the wave function $|\psi (t)\rangle $
of three spins in the $S_z$ representation $$| s_1, s_2, n \rangle,
$$ with
$$s_1=\pm 1/2, \quad s_2=\pm 1/2, \quad n=1,\ldots,52, $$
can be written as a  system of 104 coupled linear equations for the
coefficients
$$C_n (t) = \langle s_1, s_2, n | \psi (t) \rangle,$$

\begin{equation}
\begin{array}{lll}
\displaystyle
i \frac{d C_{n}}{dt}  &= B_{3} \left(\displaystyle  \frac{53}{2} - n \right)C_{n} +&\\
&&\\
&\displaystyle \frac{1}{2} B_{rf} \ \sqrt {n(52 -
n) }\ e^{i\omega t} \ C_{n + 1} +&\\
&&\\
& \sqrt{(n - 1)(53 - n)} \ e^{ -
i\omega t} \ C_{n - 1}.  \\
\label{eq4}
\end{array}
\end{equation}

Here we have used the well known expressions for the matrix elements of the
spin. In particular, for the auxiliary spin $S_{3}$ we have~:

\begin{equation}
\begin{array}{lll}
 &\langle s_1, s_2, n | S_{3}^z | s_1, s_2, n \rangle   =\displaystyle \frac{53}{2}  - n, &\\
&&\\
&\langle s_1, s_2, n  |S_{3}^+ | s_1, s_2, n + 1 \rangle  =    \sqrt{n(52 - n)}, &\\
&&\\
&\langle s_1, s_2, n | S_3^- |s_1, s_2, n-1\rangle =  \sqrt{n(52 - n)}. &\\
&&\\
&&
\label{solite}
\end{array}
\end{equation}

In Fig.~\ref{figdue} we show
 the results of our computer simulations, which confirm the
expected dynamics of the spin system. Namely, if  the control spin
$S_{1}$ is initially ``up'', the target spin $S_{2}$ does not change
its state. (See Fig. 2a.) If  the control spin $S_{1}$ is initially
``down'', the target spin $S_{2}$  changes its state from ``up" to
``down''. (See Fig. 2b.) During this operation, both spins remain in
their ground states.

In conclusion, we propose a scheme, which in effect relates
adiabatic quantum computation with traditional gate quantum
computation. Our scheme implements the change of state of the target
spin controlled by the state of the control spin in such a way that
that both spins remain in their ground states. This result is
achieved using an auxiliary spin, which mediates the interaction
between the control and target spins.

Note, that our operation can be considered as a two-bit digital
logic gate $G$ which changes the state of the target bit if and only
if it is different from the state of the control bit. We may
implement this gate using a linear polarized \emph{rf} field which
is a superposition of two circularly polarized \emph{rf} fields.
Indeed, if we assign the value ``0" to spin ``up" and the value ``1"
to spin ``down", and request that all  spins are initially in their
ground states, we will get the transformation: $G(00) = (00)$,
$G(10) = (11)$, $G(11) = (11)$, and $G(01) = (00)$. In future we
plan to study the opportunities to implement quantum logic gates
holding the qubits in their ground states.

The work by G. P. B. is supported by the NNSA of the U. S. DOE at LANL under
Contract No. DE-AC52-06NA25396.

\end{document}